# CARACTERISATION DE REACTEURS PHOTOCATALYTIQUES UTILISES POUR LE TRAITEMENT DE L'AIR

M. Faure [1,2], M. Furman [1], S. Corbel [1], MC. Carré [1], F. Gérardin [2], O. Zahraa [1]

[1] Département de Chimie Physique des Réactions – UMR 7630 CNRS-INPL, Nancy – Université, ENSIC - 1 rue Grandville, B.P. 20451, F-54001 Nancy
[2] Institut National de Recherche et de Sécurité- Laboratoire PROCEP, Département Ingénierie des Procédés – 20 avenue de Bourgogne, B.P. 27, F-54501 Vandoeuvre

**Résumé.** La photocatalyse est une technique très employée pour épurer l'air. L'étude de la dégradation photocatalytique de certains Composés Organiques Volatils (COVs) est réalisée dans des dispositifs de différents types. Afin d'optimiser les rendements de dégradation il est nécessaire de caractériser le réacteur. La première étape consiste à identifier son mode d'écoulement (réacteur pison / réacteur parfaitement agité) par détermination de la distribution du temps de séjour. Ensuite une étude cinétique peut être réalisée en fonction de différents paramètres contrôlant la réaction (temps de passage, humidité relative …). A l'issue de ces expérimentations, toutes les données sont rassemblées pour permettre un dimensionnement du système.

## INTRODUCTION

La contamination de l'air par de nombreux Composés Organiques Volatils (COVs) est un problème de santé publique du fait des effets toxiques sur l'organisme humain, même à très faible concentration. A l'extérieur, la pollution est causée par les transports et l'industrie (rejets toxiques, de solvants...). A l'intérieur, la présence de COVs est due à des matériaux et à l'usage de produits (peinture, nettoyage ...). Parmi les techniques proposées pour la réduction de la pollution de l'air intérieur et extérieur, la photocatalyse [1] se présente comme un procédé d'oxydation avancée ayant l'avantage de minéraliser totalement la plupart des COVs, dans certaines conditions.

Divers photoréacteurs (à lit fluidisé [2], à lit fixe [3], à nid d'abeille, à tube de verre imprégné) ont été construits et testés pour améliorer leur efficacité. Cependant un travail important reste à faire pour améliorer l'élaboration de ces photoréacteurs tant pour augmenter leur efficacité que pour réduire leur encombrement afin de les incorporer par exemple dans des systèmes de ventilation. Le but de ce travail est de montrer en détail la démarche à suivre pour caractériser les réacteurs photocatalytiques.

## CARACTERISATION DES REACTEURS

### 1. Généralités

Deux approches ont été développées pour décrire les écoulements dans les réacteurs [4], [5]. D'une part, une *approche fondée sur les principes de la physique théorique* avec les équations de Navier Stokes par exemple, qui vise à décrire la position, la vitesse et la pression d'un élément fluide à tout instant et en tout point de l'espace. D'autre part, une *approche systémique* qui cherche à décrire cet écoulement par des informations plus globales mais indispensables à la description des comportements transitoires des solutés réactifs que le fluide en écoulement transporte. La première approche n'est possible que pour des systèmes de géométrie simple et connue. De plus son développement demande des moyens de calculs importants. Quant à l'approche systémique, elle nécessite de définir a priori le degré de complexité désiré de l'écoulement. Elle est largement utilisée en génie chimique. L'objectif principal est de déterminer sans perturber ni détruire le système, les grandeurs globales telles que le temps pendant lequel un élément de fluide va rester dans le système par exemple.

## 2. Notion de Distribution des Temps de Séjour (DTS)

2.1 Définition

La description qui va suivre se limite à des fluides incompressibles en écoulement permanent et régime stationnaire.

Plusieurs hypothèses doivent être faites :

- le système étudié est un réacteur ouvert, possédant une entrée et une sortie à l'exclusion de toute fuite.
- le fluide qui le traverse est en écoulement permanent.
- l'écoulement à l'entrée et à la sortie se fait par convection forcée sans mélange en retour ni diffusion. Cette hypothèse est généralement vérifiée dans les réacteurs et colonnes de laboratoire du fait de la faible dimension des tuyaux d'entrée et de sortie.
- l'expérience doit être reproductible, c'est-à-dire que le milieu ne se modifie pas aléatoirement au cours du temps et le régime d'écoulement est stationnaire.

Si l'on injecte à l'entrée du réacteur, au temps t = 0 et instantanément, une quantité de traceur $n_0$, et que l'on suit son devenir à l'aide d'un détecteur linéaire en sortie, on observe une variation de la concentration en traceur au cours du temps C(t). La fraction de débit qui sort entre $t_s$ et $t_s+dt_s$ contient des molécules de traceur qui ont séjourné un temps $t_s$ dans le milieu. La courbe $C(t_s)$ est donc une représentation de la distribution des temps de séjour (DTS). En normant la courbe $C(t_s)$ par la surface qu'elle délimite, on obtient ainsi la distribution des temps de séjour $E(t_s)$ au sens de la théorie des distributions. On a alors la relation :

$$E(t_s) = \frac{C(t)}{\int_0^\infty C(t)dt} \quad \text{et} \quad \int_0^\infty E(t_s)dt_s = 1 \qquad (1)$$

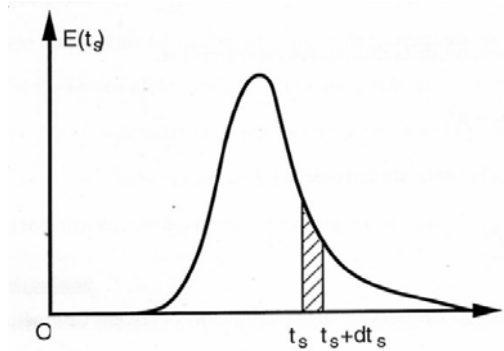

*Figure 1 : Distribution des temps de séjour [5]*

2.2 Détermination expérimentale de la DTS au moyen d'un traceur

La méthode du traceur pour déterminer la DTS consiste à associer aux molécules du fluide une proportion de molécules discernables des autres molécules par un détecteur. Cependant ce traceur doit posséder les mêmes propriétés d'écoulement. Il ne doit en particulier pas subir d'interaction avec le milieu. La technique consiste à réaliser une perturbation de concentration sur les molécules marquées à l'entrée du système et à suivre les variations de la concentration en sortie.

Plusieurs types d'injection sont possibles :

- *Injection-impulsion*

On obtient directement la fonction $E(t_s)$ en injectant le traceur instantanément à l'entrée du système, c'est-à-dire en imposant un signal impulsion (delta de Dirac) au système. Expérimentalement, on réalise un tel signal de concentration en injectant une quantité $n_0$ de traceur pendant un temps très bref (vis-à-vis d'une estimation du temps de séjour moyen dans le système (< 1 %). Cette injection doit se faire sans perturber l'écoulement. En particulier le traceur ne doit pas être injecté en trop forte concentration pour éviter la formation de courants de convection naturelle transitoires résultant d'un écart de densité entre le traceur et le fluide.

- *Injection-échelon*

Du fait de la sensibilité des détecteurs, il est souvent délicat de mesurer la réponse à une injection-impulsion. On est alors amené à réaliser une injection-échelon de traceur en faisant passer brusquement la concentration de ce dernier d'une valeur initiale $C_i$ à une valeur finale $C_f$. Le plus souvent $C_i = 0$ et on pose $C_f = C_0$. La courbe donnant la fraction de fluide marqué $C/C_0$ en fonction du temps s'appelle la courbe F. Le passage brusque de la concentration $C_0$ à la concentration nulle constitue une purge échelon et donne une courbe qui se superpose avec la précédente par simple retournement.

2.3 DTS des réacteurs idéaux

- *Réacteur piston*

Dans un réacteur piston, le fluide avance en bloc sans se mélanger. La bouffée de traceur ressort donc au bout d'un temps $t = \tau$ (le temps de passage dans le réacteur piston) sous la forme d'une impulsion étroite.

$$E(t_s) = \delta(t_s - \tau) \qquad (2)$$

- *Réacteur parfaitement agité*

Dans un réacteur parfaitement agité le traceur se répartit instantanément dans toute la masse, puis il est graduellement élué suivant une décroissance exponentielle.
La réponse à une injection-impulsion est :

$$E(t_s) = \frac{1}{\tau} \exp(\frac{-t_s}{\tau}) \qquad (3)$$

Le temps de séjour ne se marque par aucune particularité sur la courbe.

- *Réacteur quelconque*

On obtient des DTS intermédiaires entre les deux cas limites précédents.

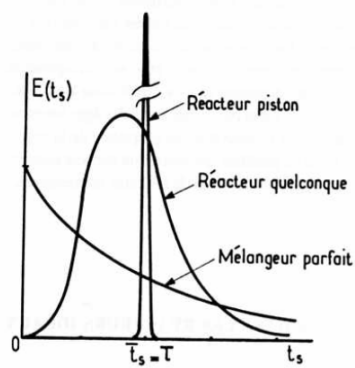

*Figure 2 : DTS des réacteurs idéaux [4]*

## 3. Modélisation des écoulements non idéaux

Après avoir obtenu une courbe expérimentale de DTS par la méthode des traceurs, on se pose généralement le problème de rendre compte de l'allure observée au moyen d'un modèle hydrodynamique simple, utilisable commodément pour prédire la conversion ou pour servir de base à l'extrapolation du réacteur. Globalement il existe deux modèles types.

### 3.1 Modèle des mélangeurs en cascade

La cascade de réacteurs agités permet de réaliser la transition entre les performances d'un réacteur agité unique et celles d'un réacteur piston. On peut tenter de représenter l'écoulement du fluide dans un réacteur réel en assimilant celui-ci à une cascade de J réacteurs agités en série de même volume total.

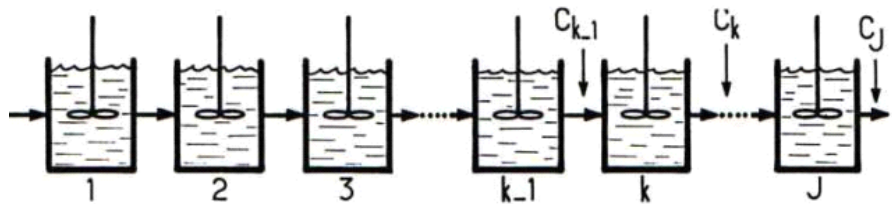

*Figure 3 : Mélangeurs en cascade [4]*

A partir d'un bilan matière sur un élément k et après diverses opérations mathématiques (transformée de Laplace, par exemple), on obtient l'expression de la DTS suivante :

$$E(t_s) = \left(\frac{J}{\tau}\right)^J \frac{t_s^{J-1} \exp(-Jt_s/\tau)}{(J-1)!} \quad (4)$$

La moyenne et la variance sont : $\quad t_s = \tau \quad et \quad \sigma^2 = (\bar{t}_s)^2 / J \quad (5)$

$$\text{Soit : } \sigma^2 = \tau^2 / J$$

### 3.2 Faible écart à l'écoulement piston

Ce modèle repose sur la superposition d'un écoulement piston convectif de vitesse u et d'une dispersion aléatoire obéissant formellement à la loi de Fick. L'expression de la DTS est donnée par :

$$E(t_s) = \frac{c}{\tau c_0} = \frac{1}{2}\left(\frac{P}{\pi \tau t_s}\right)^{1/2} \exp\left(-\frac{P(\tau - t_s)^2}{4\tau t_s}\right)$$

Où P est le nombre de Péclet, P= uL/D avec u la vitesse de l'écoulement piston convectif, L la longueur du réacteur et D le coefficient de dispersion phénoménologique.

En posant θ =$t_s$/τ on obtient l'expression suivante :

$$E = \frac{1}{2}\left(\frac{P}{\pi\theta}\right)^{1/2} \exp\left(-\frac{P(1-\theta)^2}{4\theta}\right) \quad (6)$$

## PARTIE EXPERIMENTALE

**1. Schéma du montage**
Les polluants (méthanol et éthanol) sont générés par barbotage. Afin de modifier l'hygrométrie du flux un mélange d'air sec et d'air humide en différentes proportions est réalisé. Les différents flux d'air sont ensuite mélangés et acheminés vers le réacteur photocatalytique (cf. figure 4). Les polluants sont analysés par chromatographie en phase gazeuse. Deux types de réacteurs ont été testés : un tubulaire et un annulaire.

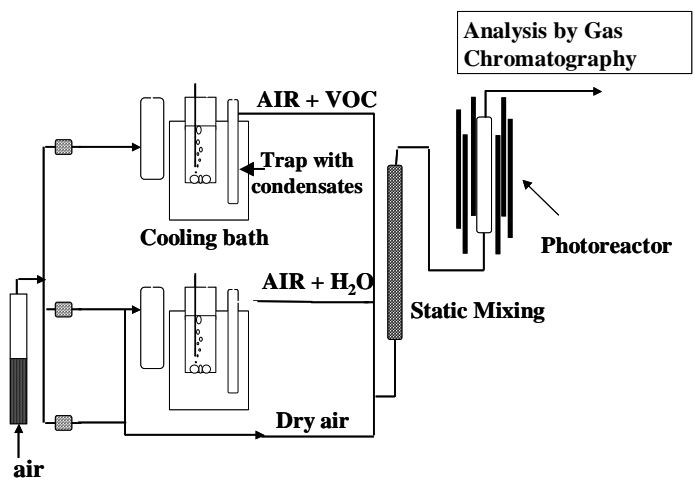

*Figure 4 : Schéma du montage*

**2. Réacteurs testés**
2.1 Réacteur tubulaire
Le premier réacteur testé est un réacteur tubulaire équipé d'un support monolithique couvert de $TiO_2$ (cf. figure 5).

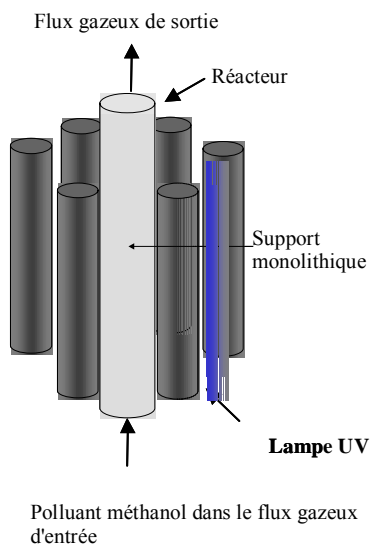

*Figure 5 : Réacteur tubulaire*

Il est entouré de six lampes UV disposées de façon à bien illuminer le catalyseur. Le support a été conçu par stéréolithographie. Il s'agit d'une méthode de prototypage rapide qui permet la fabrication d'objet solide en 3D à partir d'un modèle conçu par ordinateur. L'objet est construit couche par couche dans une cuve de monomère liquide photosensible qui durcit lorsqu'il est exposé à la lumière UV (365 nm). De nombreuses géométries peuvent ainsi être réalisées tels que des étoiles, des mélangeurs ou encore des canaux croisés (cf. figure 6). Cette technique permet notamment de faire varier des paramètres comme le diamètre des canaux ou encore le nombre de branches des étoiles.

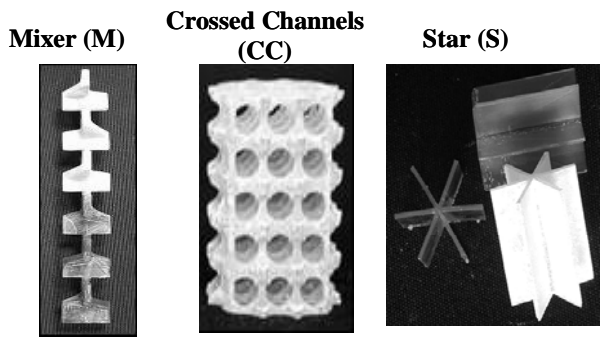

*Figure 6 : Exemples de supports monolithiques réalisés par stéréolithographie*

2.2 Réacteur à espace annulaire
Le réacteur annulaire (8 Litres) est réalisé par deux cylindres concentriques. Le cylindre intérieur est couvert d'un média photocatalytique commercialisé par la société Ahlstrom. Les lampes sont positionnées de part et d'autre du cylindre externe en pyrex. Le dispositif est également équipé de nombreux piquages de prélèvements comme le montre la figure 7.

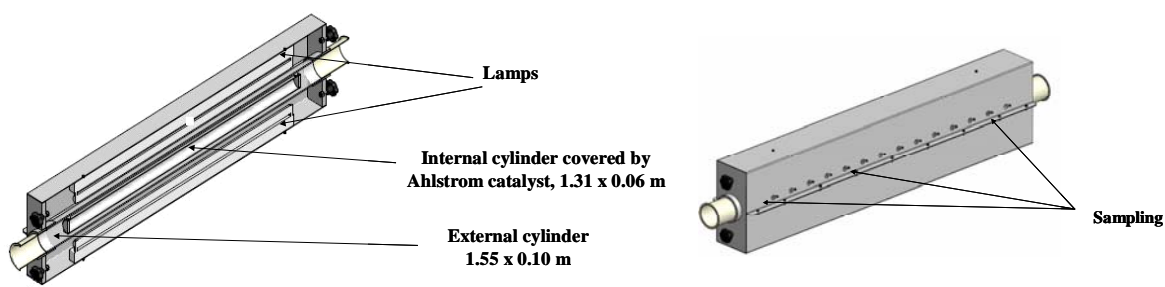

*Figure 7 : Réacteur à espace annulaire*

## MODELISATION DU PHOTOREACTEUR

### 1. Détermination de la courbe de DTS expérimentale

Afin de modéliser l'écoulement aérodynamique dans les réacteurs, des traçages à l'hélium (réacteur à espace annulaire) ou à l'hydrogène (réacteur tubulaire) ont été réalisés dans le but d'obtenir les courbes de distribution des temps de séjour pour trois temps de passage théoriques (30s, 1 minute et 1minute 30).
Concernant le traçage à l'hélium, le matériel utilisé est un spectromètre de masse (UL 100+ ®) spécifique à l'identification de l'hélium. Ce spectromètre de masse est équipé d'un dispositif de prélèvement Quicktest QT100 ® et les échantillons provenant des différents points de prélèvement sont dirigés vers la cellule d'analyse qui délivre un signal continu proportionnel à la quantité d'hélium. Le micro-ordinateur pilote l'ensemble de traçage à l'aide du logiciel PREEVENT ®. Le traçage a été fait par une injection proche d'une impulsion de traceur.

Pour le traçage à l'hydrogène, un chromatographe équipé d'un TCD a été utilisé pour réaliser les expérimentations.

**2. Détermination de la courbe de DTS théorique**

A partir des courbes de DTS expérimentales, le modèle des mélangeurs en cascade a ensuite été testé. Pour ce faire il a fallu déterminer les paramètres du modèle ($t_s$ et J) grâce à l'équation (5), et retracer les courbes de DTS du modèle à partir de l'équation (4).

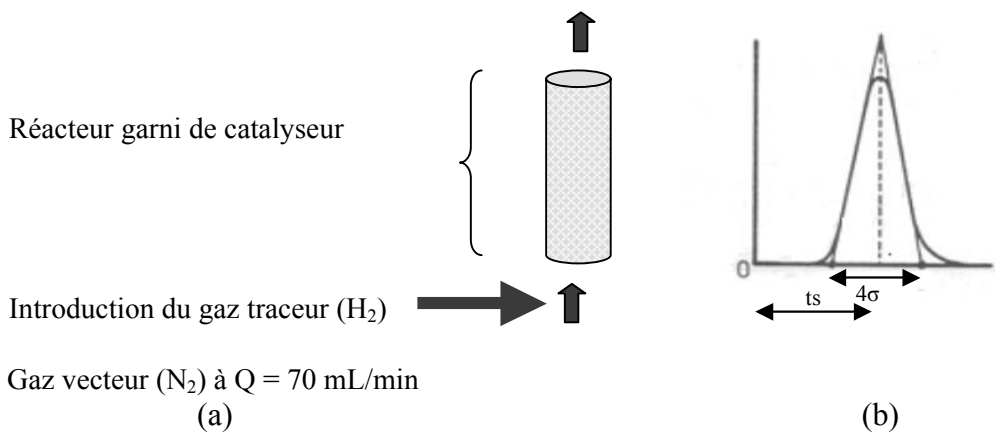

(a)            (b)

*Figure 8 : Dispositif expérimental pour le réacteur tubulaire (a) et détermination graphique de J et τ via l'équation (5) (b)*

**3. Résultats**

3.1 Réacteur annulaire

Trois traçages ont été réalisés pour chaque temps de passage. En superposant les courbes de DTS expérimentales et théoriques, on obtient des graphiques analogues à la figure 9 (b).

De plus le nombre de mélangeurs en cascade (J), pour ce réacteur, est toujours supérieur à 20 ce qui permet d'assimiler le fonctionnement du réacteur à celui d'un réacteur piston.

3.2 Réacteur tubulaire

Le résultat de DTS est représenté sur la figure 9 (a). Il s'agit des courbes obtenues pour le mélangeur statique (MS pour lequel J vaut 12), l'une des différentes structures garnissant le réacteur de longueur L'. Le tableau ci-dessous (tableau 1), donne les valeurs du paramètre J. pour toutes les structures testées. Ce dernier est peu élevé du fait essentiellement de la longueur petite des lits photocatalytiques [6].

*Tableau 1 : Valeur du paramètre J pour différentes structures garnissant le réacteur tubulaire (M=mélangeur, CCx=canaux croisés de diamètre x, Ey=étoiles composées de y branches)*

| Structure | M | Billes | CC2 | CC3 | CC4 | CC5 | CC6 | E4 | E6 | E8 | E10 |
|---|---|---|---|---|---|---|---|---|---|---|---|
| J | 12 | 20 | 14 | 14 | 10 | 11 | 11 | 9 | 10 | 9 | 13 |

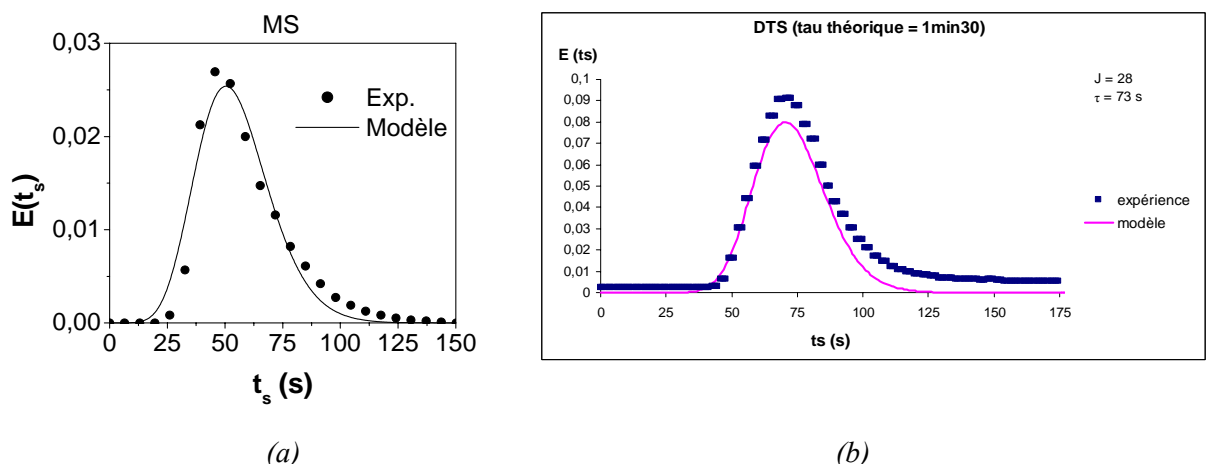

*(a)*            *(b)*

*Figure 9 : Exemple de courbe obtenue pour la DTS du réacteur tubulaire avec mélangeurs statiques (MS) (a) et du réacteur à espace annulaire (b)*

Notons par ailleurs, que le nombre J ainsi déterminé est valable uniquement dans les conditions des mesures de DTS, c'est-à-dire pour un débit gazeux Q égal à 70 mL/min. Or ce que nous recherchons, c'est une valeur de J caractérisant les lits fonctionnant à un débit plus élevé égal à 425 mL/min. L'appareillage ne permettant pas de mesurer J pour de tels débits, nous avons pris la valeur de J déterminé à 70 mL/min en l'arrondissant à l'entier le plus proche, pour faire face à ces contraintes expérimentales.

## CONCLUSION

La méthode du traçage gazeux a permis de caractériser le mode d'écoulement dans le réacteur tubulaire qui peut être ainsi assimilé à un réacteur piston Ceci est une indication indispensable pour l'établissement des bilans matière nécessaires à l'étude cinétique. Connaissant les lois cinétiques il sera alors possible de dimensionner plus précisément le réacteur dans des conditions de fonctionnement déterminées.